\documentclass{iaus}
\usepackage{graphicx}

\title[The Stellar Structures]{The Stellar Structures around Disk Galaxies}

\author[Drozdovsky et al.]{Igor Drozdovsky$^{1,2}$,
Nikolay Tikhonov$^3$,
Antonio Aparicio$^{1,4}$, Carme Gallart$^1$,
Matteo Monelli$^1$, Sebastian Hidalgo$^5$,
Edouard J. Bernard$^1$, Olga Galazutdinova$^3$ and 
the LCID team\thanks{Local Cosmology from Isolated Dwarfs; http://www.iac.es/project/LCID}}

\affiliation{$^1$Instituto de Astrof{\'{\i}sica} de Canarias,
Tenerife, Spain \break (dio@iac.es; http://www.iac.es/galeria/dio)\\[\affilskip]
$^2$Astronomical Institute of St.Petersburg State University, Russia\\[\affilskip]
$^3$Special Astrophysical Observatory, Russia
\\[\affilskip]$^4$Department of Astrophysics, University of La Laguna, Tenerife, Spain
\\[\affilskip]$^5$University of Minnesota, Twin Cities, MN, USA}

\pubyear{2006}
\volume{241}  
\pagerange{1--2}
\date{?? and in revised form ??}
\jname{Stellar Populations as Building Blocks of Galaxies}
\editors{A.~Vazdekis and R.~Peletier}

\begin{document}
\maketitle

\begin{abstract}

We present a brief summary of our current results on the stellar
distribution and population gradients of the resolved stars in
the surroundings of $\sim50$ nearby disk galaxies,
observed with space- (Hubble \& Spitzer) and ground-based telescopes
(Subaru, VLT, BTA, Palomar, CFHT, \& INT). We examine the radial (in-plane) 
and vertical (extraplanar) distributions  of resolved stars as a function of
stellar age and metallicity by tracking changes in the color-magnitude
diagram of face-on and edge-on galaxies. Our data show, that the scale
length and height of a stellar population increases with age, with
the oldest detected stellar populations identified at a large
galactocentric radius or extraplanar height, out to typically a few kpc.
In the most massive of the studied galaxies there is evidence for a
break in number density and color gradients of evolved stars, which
plausibly correspond to the thick disk and halo components of the
galaxies. The ratio of intermediate-age to old stars in the outermost fields
correlate with the gas fraction, while relative sizes of the thick-to-thin 
disks anticorrelate with galactic circular velocity. 

	  
\end{abstract}

\begin{figure}[!t]
\includegraphics[width=\columnwidth]{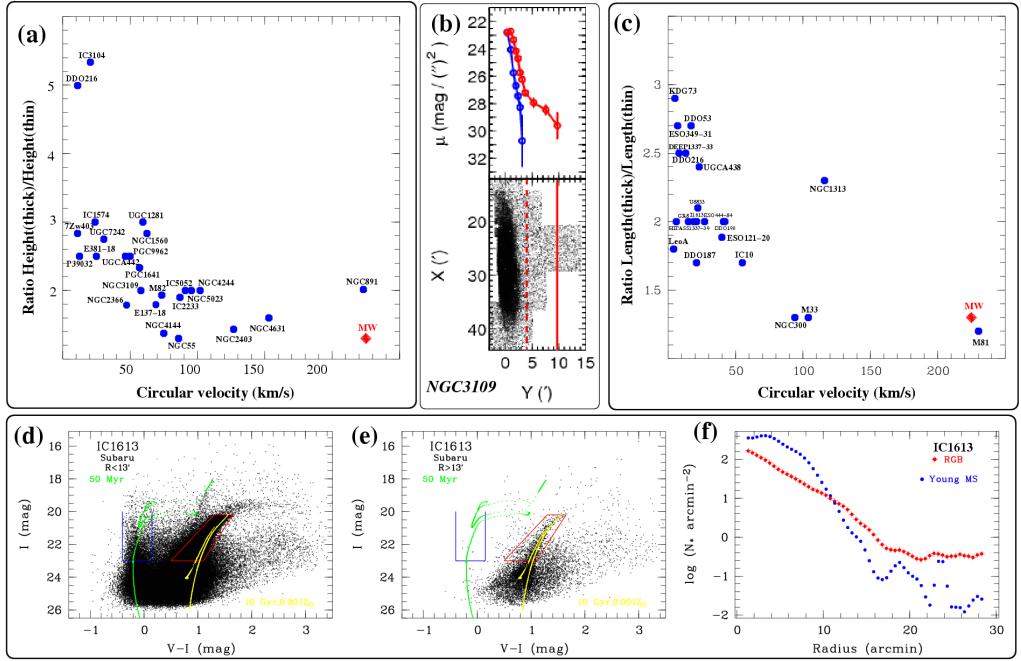}
\caption{
{\em (a)} Ratio of thick- to thin-disk height as a 
function of the rotational velocity for the high-inclination galaxies, 
based on the distribution of resolved young and evolved stars. An example 
of such study for the edge-on galaxy is shown on panel {\em (b)}. 
{\em (c))} The similar plot for the radial length ratio in the 
face-on sub-sample. 
{\em Bottom panels:} Results of our study of the stellar structures 
around IC1613, based on the data obtained with ground- (Subaru, CFHT, and INT)
and space-based telescopes (HST \& Spitzer) in collaboration with the LCID
project. {\em (d),(e)} are the CMDs for the inner and outer area, and 
{\em (f)} is a radial population gradient.  
}
\label{fig:ZvsVrot}
\end{figure}

Extended faint stellar structures have been detected around many nearby 
disk galaxies of different morphological types over the past decade, but
their nature has been a matter of some debate. 
Sometimes referred either as a 'sheet'-like (thick disk), or a spherical 
'halo' structure, these stellar formations together with streams are now
recognized as an important probe of galaxy formation and evolution
({\em e.g.\/} Abadi et al. 2006). 
%
Using the multiband photometry of individual stars,
we know that the scale length 
and height of a stellar population 
in the disk galaxies increase with age, with oldest detected 
stellar populations identified at large galactocentric radii or extraplanar
height, out to typically a few kpc ({\em e.g.\/} Drozdovsky et al. 
2003; Tikhonov, Galazutdinova \& Drozdovsky 2005; Tikhonov 2005, 2006). 
The extraplanar height of the thick
disk in low mass disk galaxies is systematically larger than the young 
thin disk of giant spirals (see Fig.~\ref{fig:ZvsVrot}a,c), suggesting that
stars in low-mass galaxies form in a thicker disk. In the most massive
of the studied galaxies (highest circular velocity), there is evidence
for a break in number density and color gradients of evolved 
Red Giant Branch (RGB) and Red Clump (RC) 
stars, which plausibly correspond to 
the transition from the thick disk to the elusive stellar spheroid, 
the halo. 

As a next step in our study we 
selected four prominent Local Group disk galaxies, NGC~6822, NGC~3109, 
IC~1613, and IC~10 to analyze in-depth the 
radial (in-plane) and vertical (extraplanar) gradients as a function of
stellar age in a wide range of galactocentric distances. 
Our data include wide-field imaging with ground-based 
telescopes aimed at reaching the old
($\sim\,10$~Gyr) stellar population, such as horizontal branch (HB)
stars --signposts of an ancient stellar population--, and deep pencil beam
fields with HST/ACS\&WFPC2 reaching the old main-sequence turnoffs (based
in part on the LCID team data). We are working on the modeling of the star 
formation history and chemical evolution of the different stellar 
sub-structures in these galaxies (see also Aparicio et al., Gallart et al., 
and Hidalgo et al., this meeting). Additionally, we are exploring the 
possibility to combine the optical data with near-infrared Spitzer/IRAC 
imaging to better disentangle the well-known age-metallicity degeneracy.   

When complemented with detailed chemical abundances and kinematic
information from the spectra of the individual stars, our optical/near-IR
photometric data will allow us to shed light on fundamental questions
about the evolution of disk galaxies, such as disk heating versus merger
scenarios and the role of these mechanisms in forming the stellar disks
and halos. 

\end{document}